\documentclass[twocolumn,prb,amsmath,amssymb,showpacs,superscriptaddress]{revtex4-1}

\usepackage{graphicx}
\usepackage{epstopdf}
\usepackage{hyperref}
\usepackage{url}
\usepackage{float}
\pdfoutput=1
\begin{document}
\newcommand{\ben}{\begin{equation}}
\newcommand{\een}{\end{equation}}

\title{Electric-field guiding of magnetic skyrmions }

\author{Pramey Upadhyaya}  
\affiliation{Department of Electrical Engineering, University of California, Los Angeles, California 90095, USA}

\author{Guoqiang Yu}
\affiliation{Department of Electrical Engineering, University of California, Los Angeles, California 90095, USA}

\author{Pedram Khalili Amiri}
\affiliation{Department of Electrical Engineering, University of California, Los Angeles, California 90095, USA}

\author{Kang L. Wang}
\affiliation{Department of Electrical Engineering, University of California, Los Angeles, California 90095, USA}

\begin{abstract}
We theoretically study equilibrium and dynamic properties of nanosized magnetic skyrmions in thin magnetic films with broken inversion symmetry, where electric field couples to magnetization via spin-orbit coupling. Based on a symmetry-based phenomenology and micromagnetic simulations we show that this electric-field coupling, via renormalizing the micromagnetic energy, modifies the equilibrium properties of the skyrmion. This change, in turn, results in a significant alteration of the current-induced skyrmion motion. Particularly, speed and direction of the skyrmion can be manipulated by designing a desired energy landscape electrically, which we describe within Thiele's analytical model and demonstrate in micromagnetic simulations including electric-field-controlled magnetic anisotropy. We additionally use this electric-field control to construct gates for controlling skyrmion motion exhibiting a transistor-like and multiplexer-like function. The proposed electric-field effect can thus provide a low energy electrical knob to extend the reach of information processing with skyrmions.
\end{abstract}
\pacs{75.78.Fg, 
75.85.+t, 
75.78.Cd, 85.75.-d}

\maketitle

\section{Introduction}

One of the central challenges faced by the field of spintronics, as an alternate to conventional charge-based information processing, is to encode information in a magnetic configuration (i.e. a collection of spins) which is both stable against thermal fluctuations and electrically manipulated with minimal energy budget. Recently, magnetic skyrmions have emerged as one such candidate configuration, opening a new sub-field of information processing with skyrmions, dubbed as \textit{skyrmionics}.\cite{Nagaosa2013} Skyrmions are topologically nontrivial magnetic textures characterized by a non-zero integer-valued skyrmion number $\mathcal{N}$, which counts the number of times the texture wraps a sphere in the spin-space\cite{rajaraman1982solitons}. Much recent interest has been fuelled by observation of the skyrmion crystal phase\cite{muhlbauer2009skyrmion,*Yu2010,*Heinze2011}, its emergent electrodynamic properties\cite{Schulz2012} and manipulation by currents at low temperatures.\cite{Jonietz2010,*Romming2013}

Meanwhile, motivated by room temperature applications, thin films with perpendicular anisotropy have been proposed as hosts for controlled nucleation and manipulation of individual skyrmions \cite{Sampaio2013}. Single skyrmions in such films can be stabilized by either Dzyaloshinskii-Moriya interaction(DMI)\cite{Dzioloshinskii, *Moriya} or dipole-dipole interaction(DDI) \footnote{In the case of DDI, more generic cylindrical magnetic domains, known as bubbles \cite{malozemoff1979magnetic}, can be stabilized, which may or may not have a non-zero skyrmion number \cite{Yu2012}. Here we focus only on textures which are topologically non-trivial and hence refer to them as skyrmions following Ref. \cite{Nagaosa2013}}competing against Zeeman and exchange energies.  
Additionally, these films can be typically sandwiched between a dielectric and a high spin-orbit coupling material resulting in broken inversion symmetry along the growth direction. Such film structures, apart from allowing for a finite DMI, have recently been shown to exhibit current-induced spin-orbit torques(SOT) \cite{MihaiMiron2010,*Liu2012}, which, in turn, result in efficient current-induced motion of chiral magnetic textures. \cite{emorinm13, *ryunnt13} These attractive properties have made such films  an ideal candidate for skyrmion-based race-track-like memory applications.\cite{Sampaio2013, *Iwasaki2013} In fact more recently, experimental evidence for skyrmionic configurations stabilized at room temperature in thin films \cite{wanjun_skyrmion, *beach_skyrmion} and multilayers \cite{fert_skyrmion} in proximity with heavy metals have been presented. In Ref. \onlinecite{wanjun_skyrmion, *beach_skyrmion}, current-induced motion has also been observed.

In this article, we propose and show|via a combination of symmetry based  phenomenology and micromagnetic simulations|that an electric-field applied across such film structures can dramatically influence current-induced skyrmion motion. In particular, we find that for a fixed current, the electric-field can control both the speed and the direction of skyrmion motion. More importantly, we show that by applying electric-fields in a certain pattern, defined via gated structures, skyrmions can be guided along desired trajectories. We also identify the mechanism for this electric-field control of skyrmion motion, which can be explained in terms of the change in the static properties of skyrmions, brought about by the electric-field induced modification of the micromagnetic free energy. Finally, utilizing this additional electrical knob we show that the realm of \textit{skyrmionics} can be extended  beyond the race-track-like memory applications, to possibly realize logic and computing functionalities.

\begin{figure}[h]
\centering
\includegraphics[width=0.5\textwidth]{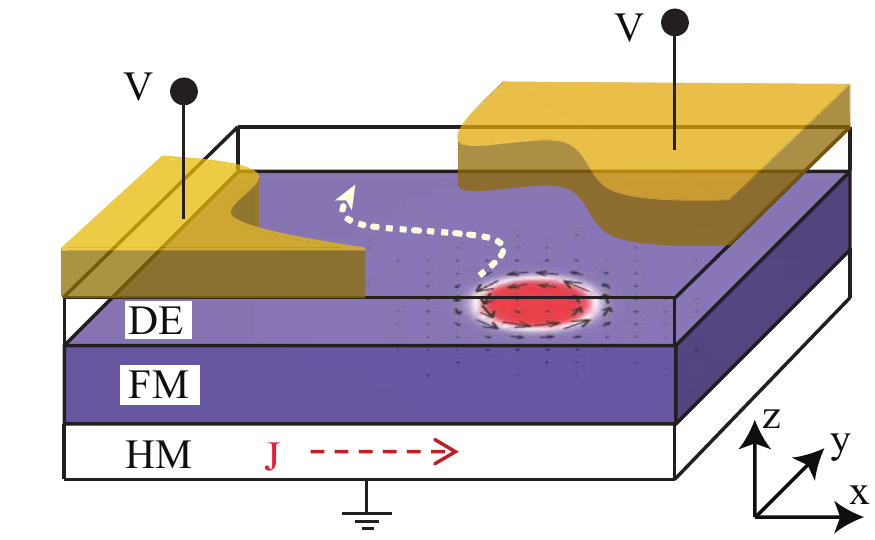}
\caption{Schematic of the system: A ferromagnet(FM) layer with a skyrmion, sandwiched between a heavy metal(HM) and a dielectric, breaking the symmetry along the $z$ axis. A current through HM with density $J$ drives skyrmion via SOT, while simultaneously local electric fields (oriented along $z$) can be applied at a gate region by an external voltage $V$.}
\label{schematic}
\end{figure}

\section{Electric-field coupling: Phenomenology}
We begin by constructing a general phenomenological model describing electric-field coupling to the magnetization dynamics based on the symmetries of the structure. We are interested here in two dimensional ferromagnetic films considered to have the following symmetry properties (see Fig.~\ref{schematic}): broken inversion symmetry along the growth axis (oriented along $z$), by interfacing it with different materials on both sides and application of an electric field, which is assumed to be applied along the $z$ axis via gates. While for simplicity, the structure is considered isotropic about the $z$ axis \footnote{As drawn, the isotropicity about the $z$ axis is broken \textit{globally} by the gates, however, here we are interested in electric-field coupling terms which result in local modification of magnetization dynamics.}. In addition, we are interested in the temperature regime well below the Curie temperature of the ferromagnet, where, suppressing the magnitude fluctuations of magnetization, the magnetization dynamics can be expressed within the Landau-Lifshitz-Gilbert framework as\cite{landauBK80} : 
\ben
\partial_t{\bf m}=-\gamma {\bf m \times H_*}+\alpha{\bf m \times \partial_t m}.
\label{LLG2}
\een
Here ${\bf m}$ is a unit vector along the magnetization ${\bf M}$ (with magnitude $M_s$), i.e. ${\bf M} \equiv M_s{\bf m}$, while $\gamma$ and $\alpha$ label gyromagnetic ratio and Gilbert damping parameters, respectively. Moreover, we have defined ${\bf H_*}\equiv {\bf -\delta_M\mathcal{F}+H}(J,E)$, with ${\bf -\delta_M\mathcal{F}}$ representing the effective field contribution derivable from a micromagnetic free energy $\mathcal{F}$, and $H(J,E)$ denoting the contribution due to application of current(having a current density, $J$) and electric-field($E$). In its minimalistic form, the micromagnetic free energy is comprised of  the classical dipole-dipole interaction $\mathcal{F}_d$, the exchange interaction parameterized by an exchange constant $A$, and the Zeeman energy due to an applied magnetic field ${\bf H_a}$. In addition, the breaking of inversion symmetry about the $z$ axis in the presence of spin-orbit interaction gives rise to interfacial perpendicular anisotropy and DMI\cite{Dzioloshinskii, *Moriya}, whose strengths are parameterized by $K$ and $D$, respectively, resulting in:

\begin{align}
\mathcal{F}=\mathcal{F}_d +A[(\partial_x {\bf m})^2 + (\partial_y {\bf m})^2]- {\bf M\cdot H_a} \\ \nonumber - Km_z^2 + Dm_z{\bf \nabla \cdot m}.
\label{free}
\end{align}
Here, ${\bf \nabla}\equiv \partial_x{\bf x}+\partial_y{\bf y}$. Similarly, the current and electric field-induced spin-orbit fields, obtained upon imposing the above mentioned structural symmetries on Eq.~\ref{LLG2} (noting  $J$ and $m_z$ are even, while $E$, $m_x$ and $m_y$ are odd under inversion about $z$ axis) and, in the spirit of linear response, retaining terms linear in $E$ and $J$, are given by:

\begin{align}
{\bf H}(J,E)=\eta_D J{\bf m \times y} +  \zeta_D Em_z{\bf m \times z} \\ \nonumber +\eta_F J{\bf y} + \zeta_F Em_z{\bf z}  .
\label{field}
\end{align}
The terms proportional to $\eta_D$ and $\zeta_D$ represent the dissipative components, i.e. which cannot be derived from a free energy, while terms proportional to $\eta_F$ and $\zeta_F$ represent current and electric-field-induced modification of $\mathcal{F}$. In particular, the terms proportional to current density represent the field-like and anti-damping like components of spin-orbit fields, microscopically derived from spin-Hall and Rashba/Edelstein effects\cite{PhysRevLett.83.1834, *rashba, *Edelstein}. Additional current-dependent terms representing corrections due to angle-dependence of spin-orbit field and symmetry-allowed dissipative torques inexpressible in the simple anti-damping like form \cite{Garello2013, *Yu2014} are not included for simplicity. On the other hand, the parameter describing coupling of electric-field to magnetization, denoted by $\zeta_F$, can be identified as electric-field dependent perpendicular magnetic anisotropy. Microscopically, electric-charge- \cite{Duan2008,*Nakamura2008, *maruyamaNATNANO09}, strain- \cite{Eerenstein2006} and ionic movement-induced \cite{Bauer2014} anisotropy modification are few examples  which are gaining increased attention in the spintronics community for exploring, both, fundamental physics and potential low-power technological applications \cite{amiri2012voltage}. Particularly for domain-wall based devices, the electric-field dependent anisotropy has been shown to trap domain walls\cite{Bauer2013} and proposed to induce domain-wall motion\cite{PhysRevB.88.224422}. Finally, the dissipative coupling of electric-field should become relevant for very thin dielectrics \cite{Silas} which, to the best of our knowledge, has not yet been measured experimentally.

At this point we mention for completeness that terms, allowed by symmetry, where electric-field and current couple to gradients of magnetization,  which could themselves be interesting to explore in proposed experimental structures, have been omitted from Eq.~\ref{free}. For example, the electric-field could generate a term of the form $\sim Em_z{\bf \nabla \cdot M}$, describing microscopic electric-field dependence of DMI. Such electric-field-induced DMI has been proposed to induce a transverse ME effect in spiral spin magnets \cite{electricDMI}. Motivated by including minimal ingredients to stabilize skyrmions and study non-trivial effects of electric-field on their dynamics based on the available experimental data, we set $D=\eta_F=\zeta_D=0$ in the following sections. However, the qualitative features of electric-field control and its mechanism are expected to be extended for the case when some of these parameters are included as well.

\begin{figure}[h]
\centering
\includegraphics[width=0.5\textwidth]{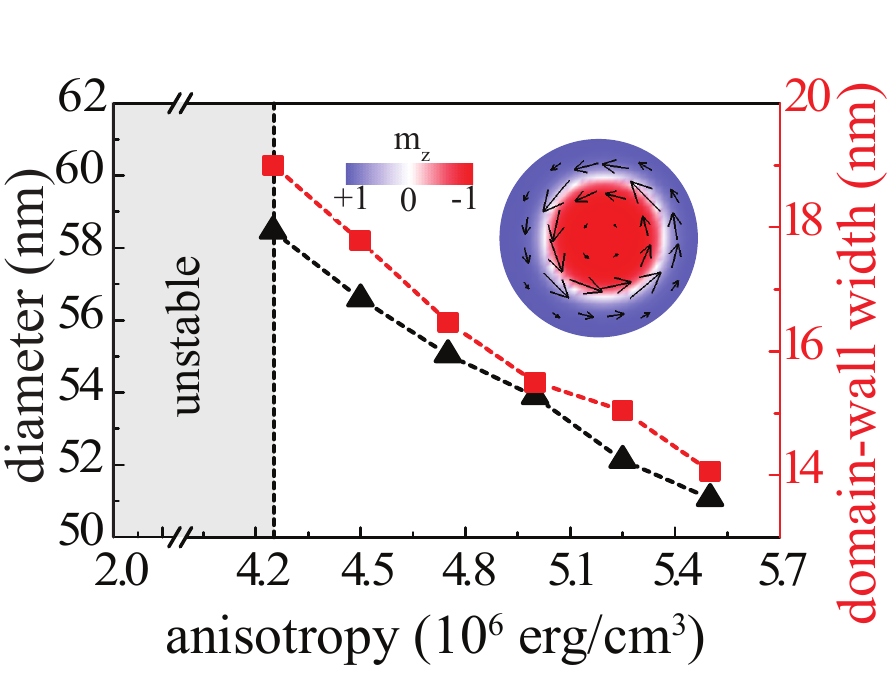}
\caption{Static properties of skyrmion as a function of perpendicular anisotropy $K$: broken vertical line marks approximate boundary between two regions. For $K>4.25\times 10^6$~erg/cm$^3$, skyrmion is stable, with its micromagnetic configuration for $K=4.5\times 10^6$~erg/cm$^3$ shown at top right corner. Here $m_z$ denotes the out of plane component of the magnetization vector, while the in-plane components are shown with an arrow. The diameter and domain-wall width surrounding the skyrmion in this region is represented by triangle and square markers, respectively. For $K<4.25\times 10^6$~erg/cm$^3$ the skyrmion is unstable, instead a multi-domain/ in-plane state is favored.}
\label{static_prop}
\end{figure}

\section{Electric-field effect: statics}
The dynamic control of skyrmions, as will be explained later, can be understood in terms of the modification of their static properties. To this end,  we discuss first the electric-field effect on the static properties of a skyrmion. In general, two separate regimes can be defined based on the stability of the skyrmion. In the first regime, the skyrmion remains stable with a smooth change in its diameter $D_s$, and the domain-wall width $\Delta_w$, surrounding the skyrmion. This change occurs in order to minimize the micromagnetic energy corresponding to the electric-field induced modification of perpendicular anisotropy: the energy cost, due to deviation of magnetization from the easy axis in the wall region, increases with increasing $K$. In an attempt to reduce this energy cost, the domain wall area ($\sim 2\pi D_s \Delta_w$) is reduced resulting in decrease of both diameter of skyrmion and the wall-width. In the second regime, on the other hand, the skyrmion itself becomes unstable. This instability occurs close to the out-of-plane to in-plane transition of magnetization when the perpendicular anisotropy is lowered to a value, such that, the perpendicular anisotropy energy is overwhelmed by the dipole-dipole interaction, due to the out of plane component of the magnetization \cite{malozemoff1979magnetic}. 

To demonstrate the qualitative picture presented above in a specific numerical model, we perform micromagnetic simulations for a DDI-stabilized skyrmion(i.e. with DMI strength set to zero) with saturation magnetization, $M_s=920$~emu/cm$^3$ and exchange stiffness $A=1$~$\mu$erg/cm. In order to take electric-field-induced modification of perpendicular anisotropy into account, the strength of perpendicular anisotropy was varied in the range $K=[2$-$5.5]\times 10^6$~erg/cm$^3$. This range corresponds to films with $K=4.25\times 10^6$~erg/cm$^3$, \cite{CoPd} in the absence of electric-field, and a maximum value of electric-field being $\sim 0.1$~MV/cm, for a coupling constant $\zeta=20$~erg/(V$\cdot$cm$^2$) \cite{Carman}. These material parameters are typical for strain-based ferromagnet/dielectric heterostructures such as CoPd(alloys)/PMN-PT, which are chosen as they exhibit one of the largest values of magneto-electric coupling (i.e. $\zeta$) at room temperature \cite{Carman}. In addition, a DDI skyrmion state has been found to be stable in finite-sized nanodisks of cobalt- and iron-based ferromagnets with similar material parameters \cite{Hehn1996,*Moutafis2007}. The simulated magnetic films are assumed to be two-dimensional with periodic boundary conditions in the film plane and a thickness of $t=32$ nm. Additionally, an external magnetic field of strength $H_{a}/H_d=0.3$ with $H_d=4\pi M_s$, is applied along the $z$ axis, which is needed to stabilize a DDI skyrmion for films of infinite extent \cite{malozemoff1979magnetic}. All micromagnetic simulations are performed using LLG Micromagnetic Simulator \cite{LLG} with temperature set to zero. 

To study the static properties of a skyrmion, the initial magnetization was set close to a film with a single skyrmion state, i.e. with a central region of sufficiently large diameter pointing along $-z$ and the rest of the film oriented along $+z$. This initial configuration was then allowed to relax to equilibrium for each value of perpendicular anisotropy. The result of such a procedure is summarized in Fig.~\ref{static_prop}. For anisotropy strength $K<K_c=4.25\times 10^6$~erg/cm$^3$, the single skyrmion state is unstable, favoring a multi-domain (for relatively strong $K$, i.e $3<K<4.25\times 10^6$~erg/cm$^3$) or in-plane magnetization (for still weaker anisotropies, i.e. $K<3\times 10^6$~erg/cm$^3$). Such a behavior is typical for films, where the out-of-plane magnetization-induced dipole energy ($\sim M_s^2$) starts dominating the perpendicular anisotropy energy ($\sim K$). This picture can be further confirmed by calculating the so-called quality factor $Q\equiv K/2\pi M_s^2$ which has a value of $Q_c\sim 0.8$, i.e. close to one, at the critical anisotropy, $K_c$. On the other hand, for $K>K_c$ in the simulations, the resulting equilibrium configuration is that of a single skyrmion (shown in the inset of Fig.~\ref{static_prop})\footnote{A second transition corresponding to the collapse of skyrmion is expected to occur on reaching a critical radius for higher anisotropies as observed for bubbles \citep{malozemoff1979magnetic}. To see that transition in micromagnetic simulations finer grid and higher anisotropy values need to be explored }. As expected, in order to minimize the dipole-dipole interaction energy in the wall, the DDI skyrmion is ``Bloch-like" with skyrmion number $\mathcal{N}=1$. Moreover, increasing anisotropy results in decreasing diameter and wall-width surrounding the skyrmion, supporting the qualitative argument presented above. 

\section{Electric-field effect: dynamics}
In this section we show how the modification of static properties of the skyrmion results in the electric-field control of current-induced skyrmion motion. Corresponding to the two regimes, defined for modification of static properties, there exist two qualitatively different regimes for the dynamic control.(i) When a uniform electric field with a magnitude below a critical value is applied, such that the skyrmion remains stable, the change in $D_s$ and $\Delta_w$ results in change in the magnitude and the direction of skyrmion motion. (ii) On the other hand, defining unstable regions for skyrmions by application of a non-uniform electric-field above the critical value, results in constraining skyrmion motion along desired paths (which we refer to as the ``guiding regime" below). To gain more insight into this connection between electric-field-induced static and dynamic control, we begin with an analytical model of current-induced skyrmion motion based on Thiele's collective coordinate approach \cite{Thiele}. 

\begin{figure}[h]
\centering
\includegraphics[width=0.5\textwidth]{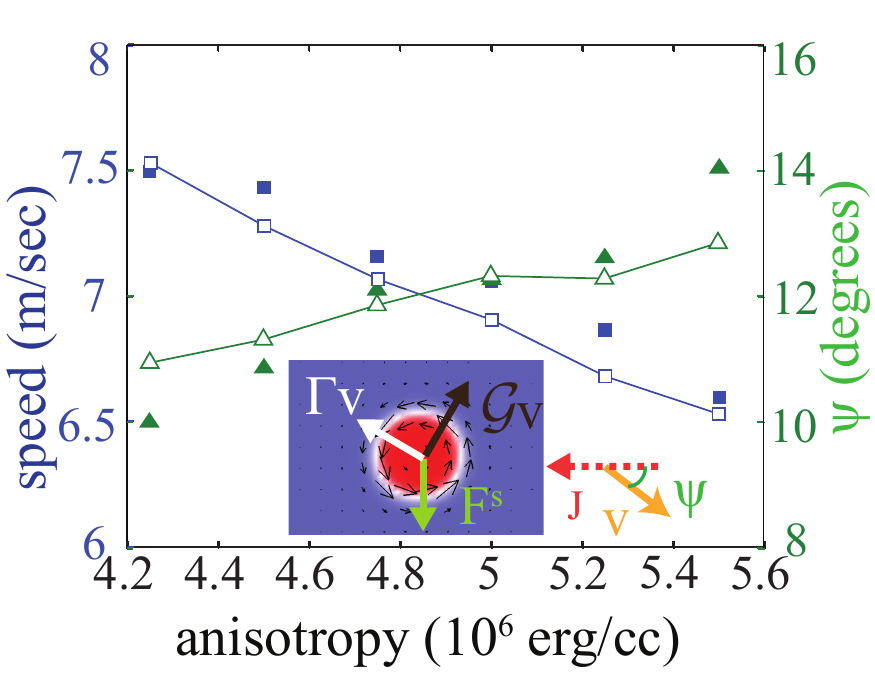}
\caption{Velocity modulation for uniform electric-field: the schematic in the bottom shows the steady-state of skyrmion moving with velocity ${\bf v}$ and the corresponding forces acting on it. ${\bf F^s}$ represents current-induced spin-orbit force, while $\mathcal{G}{\bf v}$ and $\Gamma {\bf v}$ represent the Lorentz and damping force, respectively. Open square and triangle markers represent speed and the angle $\psi$ (between current and {\bf v} as shown in the schematic) as obtained from analytical model. The corresponding speed and angle from micromagnetic simulations are represented by the closed square and triangle markers.  }
\label{vel_comp}
\end{figure}

\textit{Analytical model}| 
In the collective coordinate approach, the relevant degrees of freedom, describing the equation of motion of a magnetic-texture, are given by the so- called soft modes \cite{Bazaliy2008}. The most relevant soft modes for a skyrmion in an infinite film are its rigid translations in the film plane. Consequently, the collective coordinates included for describing the motion of a skyrmion are $(X,Y)$, which label the position of the skyrmion's center along the $x$ and $y$ axis, respectively. Within this approximation, LLG Eq.~\ref{LLG2} can be reduced to the following equation of motion \cite{Thiele,*Bazaliy2008,*Sampaio2013}:
\ben
{\bf G \times v}- \Gamma {\bf v}+{\bf F^s}-{\bf F^u}=0,
\label{EOM2}
\een
where ${\bf v}\equiv(\dot{X},\dot{Y})$ is the velocity vector, while ${\bf G}\equiv (0,0,\mathcal{G})$ with $\mathcal{G}=\int {\bf m} \cdot(\partial_x {\bf m} \times \partial_y{\bf m})$. Inclusion of other modes, such as wall fluctuations, results in deviations from the rigid motion approximation and endows Eq.~\ref{EOM2} with a ``mass term" \cite{makhfudz2012inertia}, which is not the focus of the current study. The first term describes the ``Lorentz force" \footnote{strictly speaking the dimension of each term in Eq.~\ref{EOM2} is that of a velocity but for the ease of discussion they are referred here as forces}due to the skyrmion's non-trivial topology induced fictitious magnetic field \cite{volovik1987linear}. Furthermore, the integrand in $\mathcal{G}$ is the local solid angle, making $\mathcal{G}$ independent of the exact magnetization texture and  simply proportional to the skyrmion number, i.e. $\mathcal{G}=4\pi \mathcal{N}$. The second and third terms originate from the Gilbert damping and current-induced spin-orbit field term in LLG Eq.~\ref{LLG2}, respectively, with $\Gamma=\alpha \int \partial_{x} {\bf m} \cdot \partial_{x}{\bf m} $ and $F_i^s=\gamma \eta J\int \partial_i{\bf m}\cdot({\bf m \times y})$, where $i$ labels the cartesian coordinates. Finally, the fourth term represents the force due to variation of the skyrmion's ``potential energy" $U$, i.e. ${\bf F^u}= -{\bf \nabla}{U}$. The skyrmion's potential energy is, in turn, defined as $U(X,Y)\equiv\int \mathcal{F}+\zeta_F Em_z^2/2$, where the integral is performed for the magnetic configuration of a skyrmion with its center located at $(X,Y)$. In contrast to $\mathcal{G}$; $\Gamma$, ${\bf F^s}$ and ${\bf F^u}$ depend on the exact functional form of the magnetization. 

In general, the functional form of ${\bf F^u}$ will depend on the interaction of the skyrmion with its ``environment" and will be discussed for the two regimes below, while analytical expressions are obtained for the dissipative and spin-orbit forces using Thiele's ansatz \cite{Thiele70} for a skyrmion's magnetic configuration as $\Gamma=\alpha \pi^2D_s/\Delta_w$ and ${\bf F^s}=(0,\gamma\eta J\pi^2D_s/2)$. Within this Thiele's ansatz, the integrand for $\Gamma$ and ${\bf F^s}$ is assumed to be non-zero near the wall region, where the magnetization profile, parameterized by the polar($\theta$) and azimuthal($\phi$) angles as ${\bf m}\equiv (\sin \theta\cos \phi, \sin \theta\sin \phi, \cos \theta)$, is given by $\cos \theta(r)=\tanh(\pi r/\Delta_w)$. Additionally, in order to compare to micromagnetics, the Bloch-like DDI skyrmion configuration with $\mathcal{N}=1$ is used, i.e. $\phi(\varphi)=\varphi + \pi/2$ with $(r,\varphi)$ representing the two-dimensional film plane in cylindrical coordinates.

\begin{figure*}
\centering
\includegraphics[width=\textwidth]{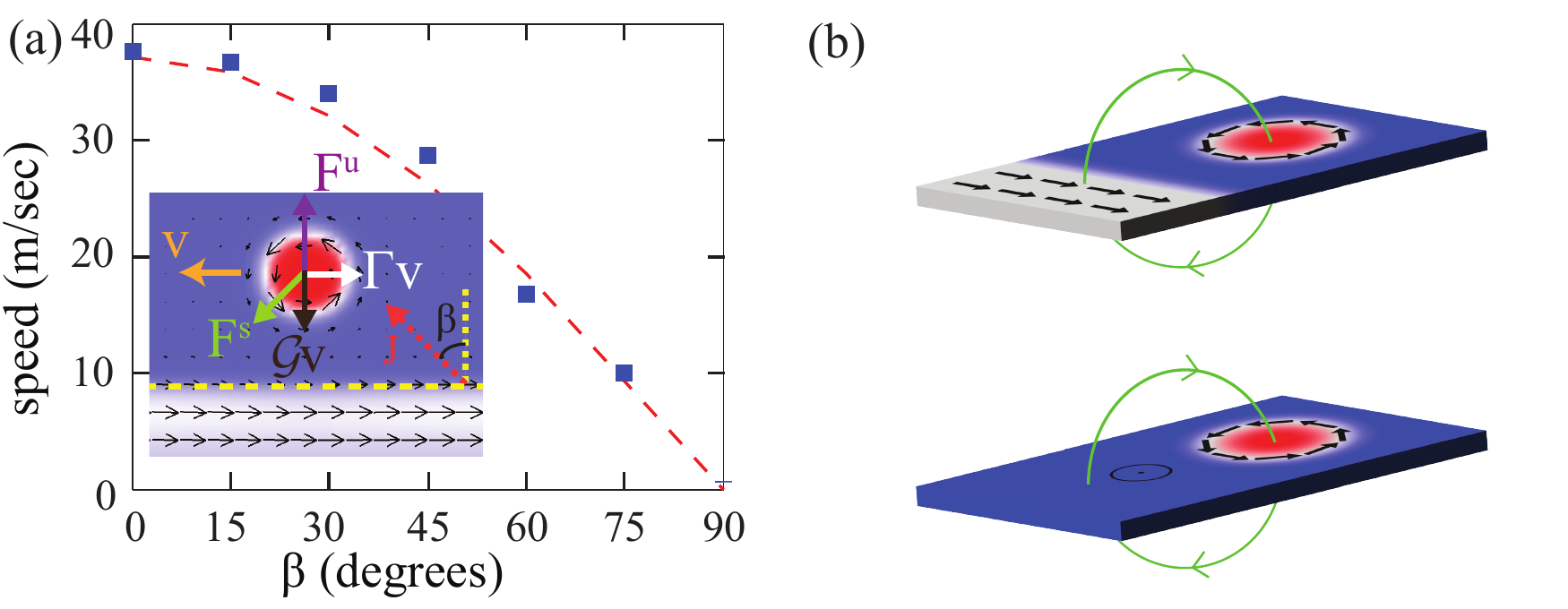}
\caption{Skyrmion guiding regime: (a) the bottom schematic shows the steady-state of a skyrmion moving along the boundary between stable and unstable regions (marked by a horizontal broken line) and corresponding forces. The angle between current and the normal to this boundary is represented by $\beta$. In addition to current-induced force, the damping force and the Lorentz force (as also shown in Fig. ~\ref{vel_comp}), there exist a dipolar force between the skyrmion and in-plane magnetization in the unstable region, denoted by ${\bf F^u}$. The skyrmion's speed along the boundary as a function of $\beta$ as obtained from the micromgnetic simulations is shown by the square marker. While the broken curve shows the result from the analytical model, i.e. Eq.~\ref{guiding_eq}, for same current density. The origin of the dipolar force is illustrated in (b), which compares the interaction of a skyrmion with the environment in the presence (top schematic) and the absence (bottom schematic) of an unstable region. The red circular region represents the core of the skyrmion, with the magnetization inside and outside the skyrmion pointing along - and + $z$, respectively, while the lines emanating from the cylinder represent the dipolar field due to the core region. The magnetization in the unstable region for the top schematic points in the plane as shown by the arrows.}
\label{guiding_comp}
\end{figure*}

We begin by looking at the regime when a uniform electric-field, having a magnitude below the critical value required to make the skyrmion unstable, is applied to the infinite film. In this case, the skyrmion's potential energy is independent of its location and hence ${\bf F^u}=0$. Consequently, solving Eq.~\ref{EOM2} for velocity, we obtain:
\ben 
v=\frac {\gamma \eta JD_s}{8\sqrt{1+(\alpha\pi D_s/4\Delta_w)^2}} ;  \psi=\tan^{-1}(\alpha \pi D_s/4\Delta_w),
\label{vel}
\een
where $v$ is the skyrmion's speed and $\psi$ is the angle between the skyrmion's motion and the current (see inset of Fig.~\ref{vel_comp} for a  schematic). This constitutes one of the main results of the analytical model, relating the skyrmion's static and dynamic properties. The electric-field control of skyrmion's motion in this regime is thus expected to enter via modification of $D_s$ and $\Delta_w$, as discussed in section III, and is shown in Fig.~\ref{vel_comp} for the diameter and domain wall width extracted from static micromagnetic simulations. These analytical results are compared later against the velocities and angles extracted directly from dynamic micromagnetic simulations.

The dependence of $U$ on electric-field suggests another qualitatively different route to control the skyrmion's dynamics via application of non-uniform electric-field and consequently ``engineering" desired forces ${\bf F^u}$. Interestingly in multiferroics, electric-field gradient-induced forces have been theoretically shown to induce Hall-like skyrmion motion\cite{PhysRevB.87.100402}. Here, we propose the scenario where unstable regions [such as the one shown in the inset of Fig.~\ref{guiding_comp}(a)] can be defined for skyrmions, where magnetization deviates from the out-of-plane direction,  due to the application of an electric field above the critical value. In this case, ${\bf F^u}$ becomes non-zero as the skyrmion approaches the boundary between the stable and unstable regions due to dipolar interaction, as explained next. The dipolar fields originating from the ``core" (i.e. the region inside the skyrmion's diameter) point in a direction opposite to the core's magnetization in the region outside the skyrmion's core. Deviation of magnetization from this dipolar field direction in the unstable region will thus result in a higher dipolar energy when compared with skyrmions interacting with the film in the absence of unstable regions [as illustrated in Fig.~\ref{guiding_comp}(b)]. Consequently, as the skyrmion moves closer to the boundary between the stable and unstable regions, the dipolar energy increases and results in ${\bf F^u}$ preventing the skyrmion to enter the unstable region. Due to this additional ${\bf F^u}$ a new steady state is expected. This situation is very similar to the case of a skyrmion driven in nanowires, where in steady state the skyrmion is driven along the nanowire, with the dipolar forces from the boundary of the nanowire balancing the Lorentz force\cite{Sampaio2013}. Thus generically it is expected that, for a range of drive currents, the skyrmion can be guided along the boundary between the stable and unstable regions, as depicted schematically in Fig.~\ref{guiding_comp}. The advantage here, as compared to the nanowire case, is that this boundary is defined by application of the electric-field and hence can be constructed on the fly, providing a controllable method to ``rewire skyrmion current". A general analytical expression for the velocity depends on the details of ${\bf F^u}$, however a simple expression can be derived for the case when the magnetization configuration in the unstable region has translational symmetry along the boundary [such as the one shown in the schematic of Fig.~\ref{guiding_comp}(a)]. This translational symmetry demands that ${\bf F^u}$ is oriented perpendicular to the boundary, whose magnitude is denoted by $F_u$. Solving for a steady state with the skyrmion moving along the boundary, taking the angle between the current and the normal to the boundary to be $\beta$, and balancing the forces along and transverse to the boundary [as depicted in the inset of Fig.~\ref{guiding_comp}(a)], we obtain for the guiding regime:
\ben
v=\frac{\gamma \eta J \Delta_w}{2\alpha}\cos \beta ; F_u=4\pi v.
\label{guiding_eq}
\een
We can see from the above expression that by orienting the electric field gate boundary with respect to the current, i.e. changing $\beta$, the skyrmion's speed can be controlled dramatically. The position of the skyrmion with respect to the boundary, on the other hand, is given by the balance of Lorentz and dipolar forces, with the skyrmion pushed closer to the boundary for larger currents.
We will now compare these analytical insights to the micromagnetic simulations and, in particular, prove the existence of the electric-field-controlled skyrmion guiding regime.

\textit{Micromagnetics}|
The micromagnetic simulations were performed for a DDI skyrmion with the same material parameters as used in the case for static simulations. In addition a current-induced spin-orbit torque, with $\eta_d=\hbar J \theta_{s}/2eM_st$ and $\theta_{s}=0.2$ \cite{Liu2012}, was turned on as the driving force for all the simulations presented in this section.\footnote{In general the effective spin-orbit field also includes a contribution of form $\sim I{\bf y}$, which cannot drive domain-walls in perpendicular magnets \cite{emorinm13, *ryunnt13} and is thus not considered here} Here $\hbar$, $e$ and $J$ are the reduced Planck's constant, charge of an electron and current density respectively. Experimentally, this can be achieved, for example, by interfacing the ferromagnet in the ferromagnet/dielectric heterostructure with a heavy metal like Ta, Pt, W etc. \footnote{In practice additional spin-torque-induced driving forces will act on skyrmion due to current flowing through the ferromagnet itself, which are neglected here for simplicity} The dynamics were calculated by integrating Eq.~\ref{LLG2} with a damping parameter $\alpha=0.05$. In Fig.~\ref{vel_comp} we first compare the results for a uniform electric-field with a magnitude below the critical field. These results are obtained for a single skyrmion moving in a film, with periodic boundary conditions along the film plane, and a fixed current density of $J=10^8$~A/cm$^2$, while simultaneously varying the anisotropy throughout the film in the stable range (as obtained in the static simulations, i.e. for $K=[4.25-5.5]\times 10^6$~erg/cm$^3$ in Fig.\ref{static_prop}). As can be seen, both the speed and the angle of the skyrmion with respect to the current direction are modulated by application of the electric-field. Moreover for the range of electric-field magnitudes studied here, the trend predicted by the analytical model fits the obtained $v$ and $\psi$ well, both qualitatively and quantitatively, substantiating the analytical picture presented above. For the material parameters used, the electric-field modulation shown in this regime is, however, relatively weak. For a much stronger control of skyrmion dynamics, we next turn towards the other regime, i.e application of a non-uniform electric-field. 

\begin{figure}[h]
\centering
\includegraphics[width=0.5\textwidth]{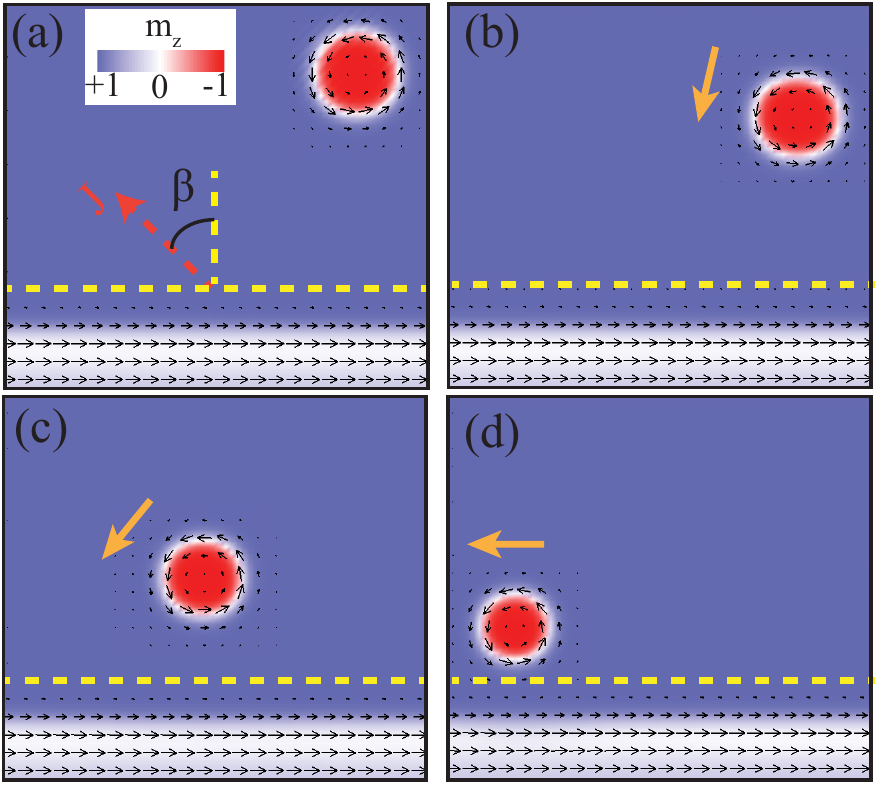}
\caption{Micromagnetic simulations showing the guiding by electric field: (a) Initial state showing a stable skyrmion with $K=4.25\times 10^6$~erg/cm$^3$ in the region above the horizontal dashed line and in-plane magnetization below it, where $K=2\times 10^6$~erg/cm$^3$. A current, with density $J$, and making an angle $\beta$ with the normal to the boundary between stable and unstable regions, provides the driving force for the skyrmion. The inset shows the color code for out of plane magnetization, while the in-plane components are shown by arrows. The snapshot of simulation after (b) 5 ns, (c) 10 ns and (d) 20 ns, are shown with the arrows alongside the skyrmion representing the instantaneous direction of the skyrmion's motion.}
\label{guiding}
\end{figure}

To demonstrate the skyrmion guiding regime by defining unstable regions for skyrmions, micromagnetic simulations were performed with the value of $K=4.25 (2)\times 10^6$~erg/cm$^3$ in the stable(unstable) regions marked in Fig.~\ref{guiding}. The dynamics was then simulated for a skyrmion driven by current, with the initial state of a single skyrmion in the stable region. In Fig.~\ref{guiding} we show various snapshots of the results of these simulations. First note that, as was discussed in the section III, the magnetization in the unstable region goes to an in-plane state. In the initial stage of the simulation, i.e. when the skyrmion is relatively far from the the unstable region [Fig ~\ref{guiding}(b)], the skyrmion behaves very similar to ${\bf F^u}=0$ case moving at an angle with respect to the current. However, as the skyrmion approaches the boundary between the stable and unstable region [Fig.~\ref{guiding}(c)] a non-zero ${\bf F^u}$ deflects the skyrmion, preventing it from entering the unstable region. After $t\sim 40$ns a new steady state is reached with the skyrmion moving along the boundary between the stable and unstable region [Fig.~\ref{guiding}(d)] and hence proving the existence of the guiding regime. The robustness of this steady state was checked by varying the magnitude of drive current density. For this example, the skyrmion is still driven along the boundary for current density below $J\sim 5 \times 10^8$~A/cm$^2$. As the drive force increases, with increasing current density, the skyrmion is pushed closer to the boundary according to Eq.~\ref{guiding_eq}. However, above $J\sim 5 \times 10^8$~A/cm$^2$ the drive force overcomes ${\bf F^u}$, resulting in the skyrmion entering the unstable region where it is annihilated. Motivated by applications, we next discuss how this regime of skyrmion guiding could provide new opportunities for constructing possible spintronic devices.

\section{Skyrmionics beyond race-tracks}
Nanosized skyrmions can be used as non-volatile information carriers in a spintronic device owing to their stability against thermal fluctuations. The information stored in the skyrmion can, in turn, be read both optically or electrically (by utilizing various magneto-resistance effects such as topological Hall effect \cite{PhysRevLett.102.186601}, tunneling- \cite{jullierePLA75} or giant- \cite{PhysRevLett.61.2472, *PhysRevB.39.4828} magnetoresistance). One such proposed example of a spintronic device is a race-track memory, where information stored in a magnetic texture is moved by current to a desired location where it can be read or written\cite{Parkin11042008}. With the discovery of current-induced skyrmion motion, skyrmions in a nanowire are considered as a promising alternate to ``conventional" domains in a race-track memory as they show very weak pinning by external impurities requiring orders of magnitude lower critical depinning currents \cite{Jonietz2010}. The new possibility of controlling ``skyrmion current" by the electrical-field presented here is thus expected to open up novel avenues of information processing with skyrmions. We demonstrate this by specifically showing via micromagnetic simulations that electric-field gates can be used to turn ``skyrmion currents" off, i.e. providing a transistor-like operation, and deflect information carried by skyrmions in specific parts of the spintronic circuit, i.e. a multiplexer-like operation.

\begin{figure}[h]
\centering
\includegraphics[width=0.5\textwidth]{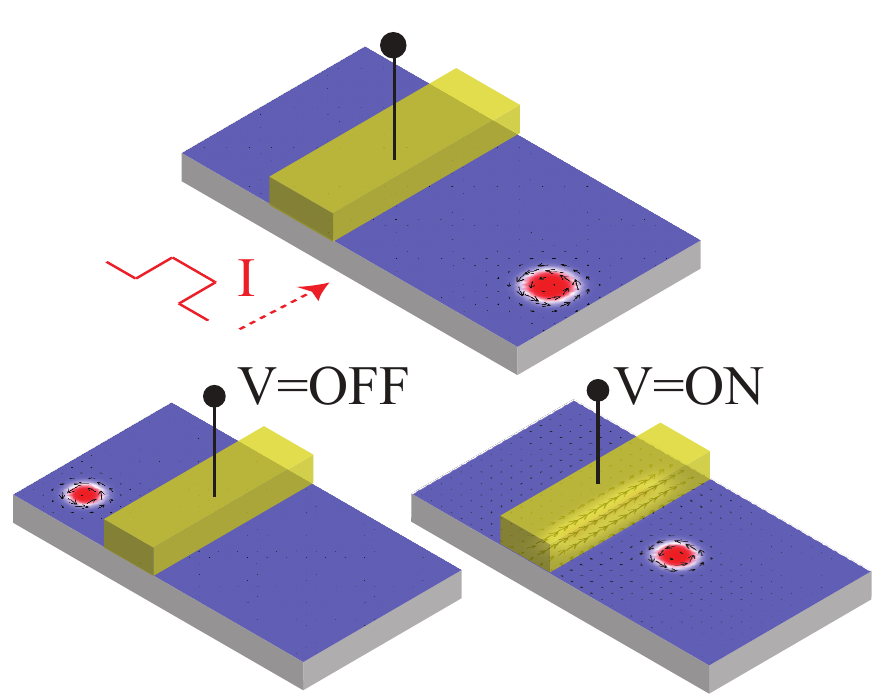}
\caption{Micromagnetic simulations showing the transistor-like function: top schematic shows the initial state with a skyrmion located on the right end of the nanowire, along with a current pulse of 10 ns passed through the heavy metal underlayer. An external electric-field can be applied locally via a voltage over a gate region indicated in the figure. The resulting location of the skyrmion at the end of the pulse when the gate is in OFF (bottom left) and ON (bottom right) state.}
\label{transistor}
\end{figure}

To demonstrate the transistor-like operation, a nanowire geometry is adopted in the simulations with width of the nanowires set at 200 nm. The anisotropy strength was $K=4.25\times 10^6$~erg/cm$^3$ throughout the simulations everywhere in the nanowire except in a region of width $50$ nm defined as the gate. To simulate the effect of electric-field applied on the gated region, the anisotropy value was dropped to $K=2\times 10^6$~erg/cm$^3$ when the gate is in the ``ON" state and kept at $K=4.25\times 10^6$~erg/cm$^3$ in the ``OFF" state. The rest of the parameters are same as for the dynamical simulations presented above. The result of the application of a drive current pulse along the $x$ axis of duration $t_p=10$~ns is shown in Fig.~\ref{transistor}. When the gate is in the OFF state, the skyrmion is driven past the gate region along the nanowire (Fig.~\ref{transistor} bottom left). On the other hand, when the same current pulse is applied in the presence of a gate voltage, the skyrmion is blocked by the gate (Fig.~\ref{transistor} bottom right), due to the repulsive dipolar forces from the unstable gated region, thus exhibiting a transistor-like function. This situation is the special case of the guiding regime, i.e. Eq.~\ref{guiding_eq}, with $\beta =90$.

\begin{figure}[h]
\centering
\includegraphics[width=0.5\textwidth]{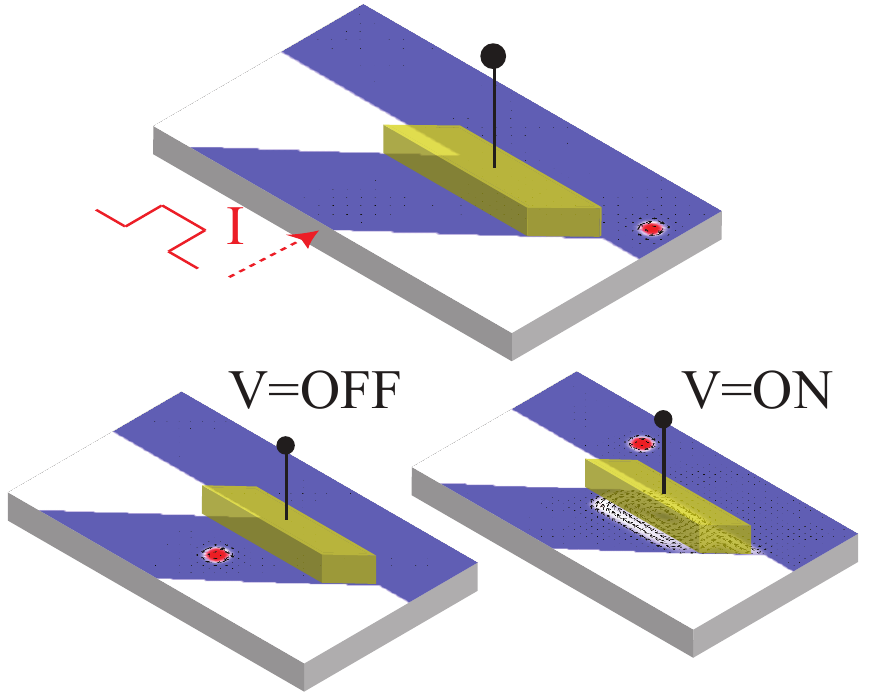}
\caption{Micromagnetic simulations showing the multiplexer-like function: top schematic shows the initial state for two magnetic nanowires coming together at the ``wing-shaped" junction with a skyrmion located at the right end of the upper arm. A local external field can be applied at the shown gate region. The result of the simulation at the end of a current pulse of 10ns passed through the heavy metal underlayer when the gate is OFF (bottom left schematic) and ON (bottom right schematic).  }
\label{multiplex}
\end{figure}

Next we simulate a ``wing-shaped" nanowire junction, with two nanowires, each of width 200 nm, meeting at an angle representing two different arms of a spintronic circuit. A gate in this case is defined on one of the arms as shown in Fig.~\ref{multiplex}, which also shows snapshots of the simulations. When the gate is OFF, a current pulse along the $x$ axis in this case drives the skyrmion to the lower arm (Fig.~\ref{multiplex} bottom left). This motion is similar to a skyrmion moving at an angle with the current due to the Lorentz force, as seen for a skyrmion moving in an infinite film. However, when the gate voltage is turned on, the skyrmion driven by the same current pulse is deflected off the gate region into the upper arm (Fig.~\ref{multiplex} bottom right). Such structures can thus be used to rewire the skyrmion current into specific parts of the circuit.

\section{Summary}

It is shown that an electric-field provides an additional degree of freedom for effective control of current-driven skyrmion motion. By using a symmetry-based phenomenology we find the general form of spin-orbit-induced electric-field coupling for thin films with broken inversion symmetry about the film plane. In general, this coupling can be classified as dissipative, i.e. not derivable from a free energy,  and conservative, i.e. terms which can be understood as electric-field- induced modification of the free energy. The conservative terms can further be identified as electric-field dependent perpendicular anisotropy (when electric field does not couple to gradients of magnetization) and electric-field dependent exchange and Dzyaloshinskii-Moriya interaction (when electric-field couples to gradient in magnetization). Motivated by applications and available experimental data, we  specifically studied effects of the electric-field dependent perpendicular ansisotropy in thin films, which also harbor a nanosized skyrmion state stabilized by dipolar interaction.  We find that there exist two regimes for controlling the static and hence the dynamical properties of skyrmions: (a) below a critical value, such that the corresponding perpendicular ansitoropy is still strong enough to overcome the out-of-plane dipolar interaction, skyrmions are a stable configuration showing a decrease in diameter and domain wall width surrounding the skyrmion with increasing anisotropy, and (b) above the critical value, the skyrmion itself becomes unstable due to formation of multi-domain/ in-plane magnetized state. The decrease in diameter and domain wall width with increasing anisotropy in regime (a) is, in turn, driven by an attempt to reduce the wall energy. This change of static properties modulates both the speed and the angle of skyrmion motion driven by a current induced spin-orbit force. This modulation can be understood within Thiele's analytical model of rigid skyrmion motion, where, the static properties enter the velocity via $D_s$ and $\Delta_w$ dependence of spin-orbit and damping forces acting on the skyrmion. On the other hand, corresponding to regime (b), desired unstable regions can be defined for skyrmions using gate voltages, which are observed in micromagnetic simulations to guide skyrmions along the boundary between stable and unstable regions.This regime of skyrmion guiding is established by appearance of additional forces due to variation of skyrmion's potential energy, caused by the dipolar interactions, as the skyrmion approaches the unstable region. Using this ability to electrically design potential energy landscapes, we demonstrate transistor-like and multiplexer-like spintronic devices controlling skyrmion current by defining electric-field  gates. The proposed electric-field control is thus expected to extend the reach of \textit{skyrmionics} into logic and computing devices.

\begin{acknowledgements}
This work was partially supported by the National Science Foundation Nanosystems Engineering Research Center for Translational Applications of Nanoscale Multiferroic Systems (TANMS), and in part by the FAME Center, one of six centers of the Semiconductor Technology Advanced Research network (STARnet), a Semiconductor Research Corporation (SRC) program sponsored by the Microelectronics advanced Research Corporation (MARCO) and the Defense Advanced Research Projects Agency (DARPA). P.U. also acknowledges partial support from the Spins and Heat in Nanoscale Electronic Systems (SHINES), an Energy Frontier Research Center funded by the U.S. Department of Energy (DOE), Office of Science, Basic Energy Sciences (BES), under Award \# DE-SC0012670. We would like to thank Yaroslav Tserkovnyak, Greg Carman, Axel Hoffman, Wanjun Jiang, and Kin Wong for useful discussions.
\end{acknowledgements}

\end{document}